\documentclass[12pt]{article}
\usepackage{times,psfrag,graphicx,theorem,epsfig,lscape,amssymb,amsmath,dcolumn, multirow,float, enumerate,latexsym,bm,subfigure}

\usepackage{comment}
\usepackage{color}
\usepackage[round]{natbib}
\usepackage{pdflscape}
\usepackage{rotating}

\newcommand{\blinding}[2]{#1}   %%% USE THIS FOR UNBLINDED
\theoremstyle{plain} 
\theoremstyle{plain} 
\theoremstyle{plain} 
\theoremstyle{plain} 
\theoremstyle{plain}

\newcommand{\bmu}{\bm{\mu}}

\newcommand{\btheta}{\bm{\theta}}

\newcommand{\bC}{\bm{C}}
\newcommand{\bQ}{\bm{Q}}
\newcommand{\bq}{\bm{q}}

\newcommand{\bV}{\bm{V}}
\newcommand{\bd}{\bm{d}}

\newcommand{\mI}{\mathcal{I}}

\newcommand{\bx}{\mathbf{x}}

\newcommand{\normal}{\mbox{N}}

\newcommand{\one}{\mathbf{1}}
\newcommand{\qed}{\hfill \mbox{\raggedright \rule{.07in}{.1in}}}

\DeclareMathOperator\bE{\mathbb E} % expectation
\DeclareMathOperator\bVar{\mathbb V} % variance

\begin{document}

\begin{center}
\vspace*{-2.5cm}

{\Large Improving Inference of Gaussian Mixtures\\ Using Auxiliary Variables}

\blinding{
\bigskip
Andrea Mercatanti \footnote{Andrea Mercatanti is researcher,
Statistics Department, Bank of Italy, Rome, Italy.
(email: mercatan@libero.it).} \quad \quad Fan Li \footnote{Fan Li
is assistant professor, Department of Statistical Science, Duke
University, Durham, NC, USA. (email: fli@stat.duke.edu).} \quad
\quad Fabrizia Mealli \footnote{Fabrizia Mealli is professor,
Department of Statistics, Informatics, Applications, University of Florence, Italy. (email:
mealli@ds.unifi.it). }

%\today
}{}
\end{center}

{\centerline{ABSTRACT}

\noindent Expanding a lower-dimensional problem to a higher-dimensional space and then projecting back is often beneficial. This article rigorously investigates this perspective in the context of finite mixture models, namely how to improve inference for mixture models by using auxiliary variables. Despite the large literature in mixture models and several empirical examples, there is no previous work that gives general theoretical justification for including auxiliary variables in mixture models, even for special cases. We provide a theoretical basis for comparing inference for mixture multivariate models with the corresponding inference for marginal univariate mixture models. Analytical results for several special cases are established. We show that the probability of correctly allocating mixture memberships and the information number for the means of the primary outcome in a bivariate model with two Gaussian mixtures are generally larger than those in each univariate model. Simulations under a range of scenarios, including misspecified models, are conducted to examine the improvement. The method is illustrated by two real applications in ecology and causal inference.

\vspace*{0.5cm}
}

\clearpage

\section{Introduction}
The idea of expanding a lower-dimensional problem to a higher-dimensional space and then projecting back has been used in statistics and other disciplines. This article discusses a specific example in the context of finite mixture models; in particular, we rigorously investigate the impact on inference for mixture models when using auxiliary variables. Finite mixture models are a large class of statistical models for studying a wide variety of practical problems; comprehensive reviews can be found in \cite{McLachlanBasford88,McLachlanPeel00}. The common idea underlying these models is that data are obtained from two or more underlying populations with common distributional form but different parameters. Formally, the data $\bx=(\bx_1,...,\bx_n)'$, where $\bx_i$ is a $m$-dimensional vector and $n$ is the sample size, follow the distribution:
\begin{equation}
f(\bx)=\sum_{k=1}^K p_k f_{k}(\bx;\btheta_k), \label{mixture_gen}
\end{equation}
with the weights $p_k$'s satisfying $p_k>0$ and $\sum_k p_k=1$. Standard choices of the densities $f_k$ include Gaussian (normal), Poisson and Student's t-distributions. Here we focus on the most widely used Gaussian mixture models.

A main source of uncertainty in estimating mixture models is attributed to the unknown mixture membership of each unit. The EM algorithm \citep{EM}, which augments the mixture membership for each unit iteratively, is the most common approach for deriving maximum likelihood estimates (MLEs) of the parameters in mixture models. For estimating the variance matrix of the ML estimator, there are three main approaches. The first involves the ``complete-data'' likelihood, where the augmented mixture membership for each unit is treated as observed \citep[e.g.][]{Louis82}. The second are resampling-based methods \citep[e.g.][]{Newton94, Basford97}. The third is based on the original ``incomplete-data'' likelihood \citep[e.g.][]{Dietz96}. An important recent work in this area is \cite{Boldea09}, who derived the analytical forms of the score vector and Hessian matrix for Gaussian mixture models with arbitrary number of components and dimension of observations. Besides the likelihood-based approaches, there is also a large literature on the Bayesian approach to mixture models \cite[e.g.][and references
therein]{West1992,West1994,RichardsonGreen97,Marin05}.

Regardless of the mode of inference used, the key to inference for mixture models is to disentangle the unknown mixtures. Our main message here is that inference for mixture models can be sharpened by jointly modeling the primary variable with available auxiliary variable(s). Despite the large literature on multivariate mixture models, cross-dimensional comparison is rare. Multivariate analysis is usually conducted when the features of several variables or the relationship between the variables is of interest, but it is seldom considered for the purpose of sharpening univariate inference. Indeed, it is not obvious why including auxiliary variables in the models would improve estimation of the parameters for the primary variable. Clearly jointly modeling the primary variable with any arbitrary random variable would in principle only increase noise. But in real applications, (auxiliary) variables are usually associated with the mixture membership. Thus, on one hand, proper utilization of those relevant auxiliary variables may provide extra information to predict the mixture membership and consequently to disentangle the mixtures. On the other hand, however, for a given sample size modeling auxiliary variables could induce extra uncertainty because of the estimation of additional parameters; further, it increases model complexity and thus the risk of mis-specification. We show that the potential benefits dominate the potential drawbacks.

There are a few empirical examples within specific contexts that display the benefit of using auxiliary variables in mixture models. In the context of causal inference, a common goal in randomized clinical trials is to evaluate the effect of a drug or a therapy on a primary clinical outcome. While measurements on other features, such as side effects, are routinely collected, they are usually analyzed separately, one at a time. When noncompliance arises, mixture models are often used since the population is heterogenous regarding compliance behavior \citep[e.g.][]{Imbens97}. \cite{Mattei13} and \cite{MealliPacini13} show that jointly modeling primary and secondary outcomes significantly sharpens the inference for the primary outcome. Another example is found in the context of small-area estimation: \cite{DeSouza92} showed that analysis based on bivariate hierarchical models, which can be viewed as a special case of mixture models, reduces the posterior standard errors of the mean small-area estimates compared to those based on univariate models. However, to our knowledge, there is no previous work that gives general theoretical justification for this practice or explains the reasons underlying the benefit of auxiliary variables in mixture models, even for special cases.

The goal of this article is to provide a theoretical basis for comparing inference for multivariate mixture models with inference for the corresponding marginal univariate mixture models, filling a gap in the literature. In particular, proceeding from the incomplete-data likelihood perspective, we will establish analytical results for several special cases, showing that multivariate analysis increases the probability of correctly allocating the mixture membership and improves precision (or equivalently, reduces standard errors) of the maximum likelihood (ML) estimates compared to the corresponding univariate analysis. Another key insight from our results, partly shown in our empirical analysis, is that the introduction of an auxiliary variable tends to regularize the model and thus reduce the prevalence and the likelihood of spurious roots. We show these benefits clearly dominate the extra uncertainty due the larger parameters set involved by the auxiliary variable. As closed-form arguments on general mixture models are difficult to obtain, our analytical derivations are focused on the simple case of bivariate mixture models with two Gaussian components; models with mis-specification (non-Gaussian) and higher dimensions are explored in simulations and real applications.

The rest of the paper is organized as follows. In Section 2, we illustrate the intuition by a simple visual example and present the main theoretical results. In Section 3, we conduct simulations to examine the small-sample comparisons between bivariate and univariate analyses under a variety of settings. Two real applications are presented in Section 4. Section 5 concludes.

\section{Comparing univariate and bivariate mixture models} \label{sec:mainresults}
\subsection{Basic setup and intuition}
Consider a mixture model of two Gaussian densities,
\begin{equation}
f(\bx)=p~f_{1}(\bx)+(1-p)~f_{2}(\bx), \label{mixture}
\end{equation}
where
$f_{k}(\bx)=\left\vert \bV_{k}\right\vert
^{-1/2}\exp \left\{ -(\bx-\bmu_{k})^{\prime }\bV%
_{k}^{-1}(\bx-\bmu_{k})/2\right\}/2\pi$ for $k=1,2$. For a univariate density, $\bx=x_1, \bmu_k=\mu_{1k}, \bV_k=\sigma_{1k}^2$, while for a bivariate density,
\begin{equation}
~\bx=\left(
\begin{array}{c}
x_{1} \\
x_{2}%
\end{array}%
\right), ~
\bmu_{k}=\left(
\begin{array}{c}
\mu _{1k} \\
\mu _{2k}%
\end{array}%
\right) ,~\bV_{k}=\left(
\begin{array}{cc}
\sigma _{1k}^{2} & \rho _{k}\sigma _{1k}\sigma _{2k} \\
\rho _{k}\sigma _{1k}\sigma _{2k} & \sigma _{2k}^{2}%
\end{array}%
\right).  \label{input}
\end{equation}
In what follows, we will use the subscript $m (=1,2)$ to denote
the outcome and $k (=1,2)$ to denote the mixture component.

\begin{figure}[!ht]
\centering
\includegraphics[scale=0.75]{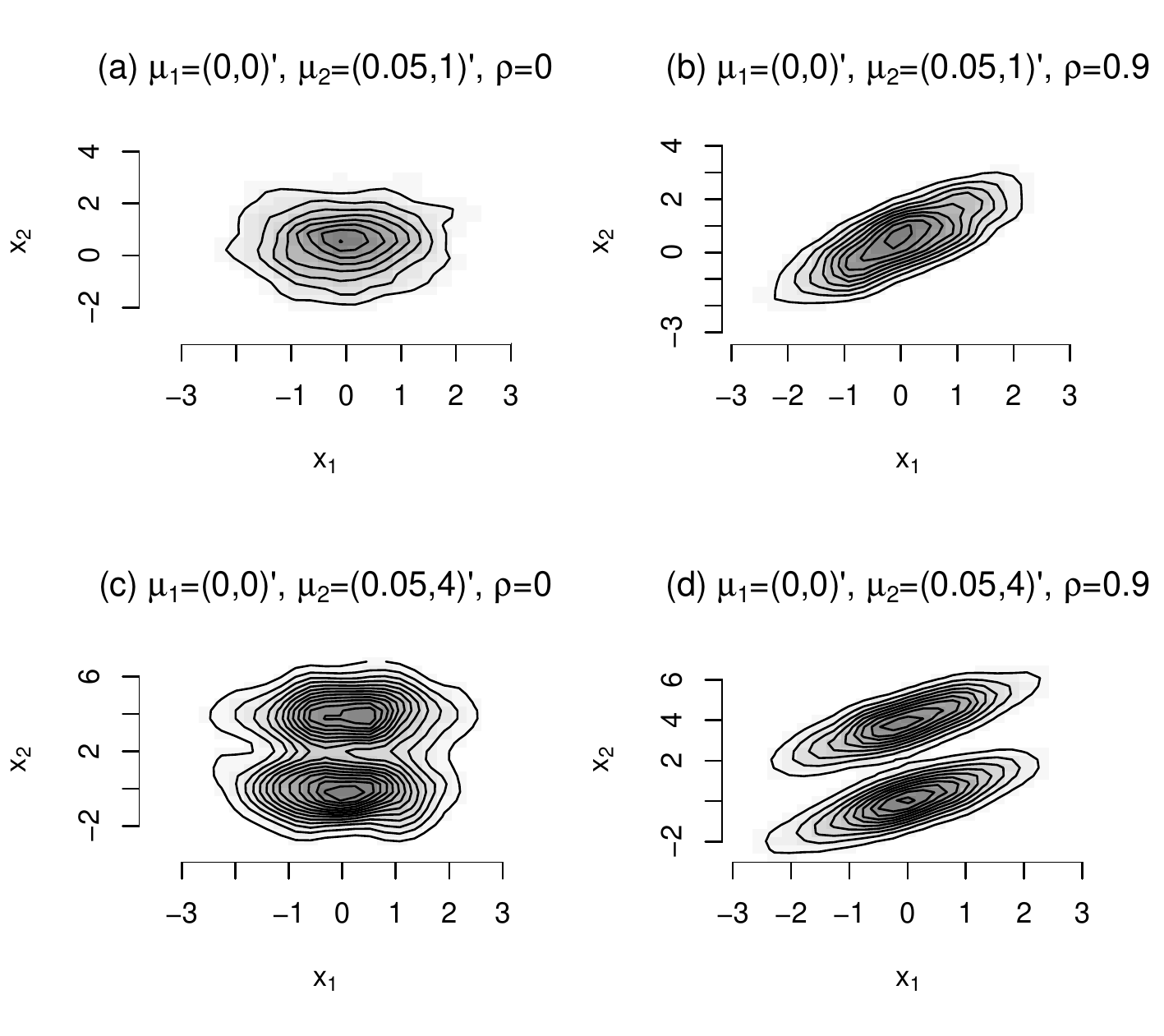}
\vspace*{-1cm}
\caption{Empirical contour plots from 1000 samples simulated from density \eqref{input}, all with $p=0.5$,
$\bV_1=\bV_2$ and $\sigma_1=\sigma_2 =1$.}
\label{contour}
\end{figure}

The intuition of the benefit of using the second outcome can be
illustrated by a simple example in Figure \ref{contour}. Consider
four sets of parameters in \eqref{input}, all with $p=0.5$,
$\bV_1=\bV_2$ and $\sigma_1=\sigma_2 =1$, but different means and
correlations: (a) $\bmu_1=(0,0)', \bmu_2=(0.05,1)', \rho=0$; (b)
$\bmu_1=(0,0)', \bmu_2=(0.05,1)', \rho=0.9$; (c) $\bmu_1=(0,0)',
\bmu_2=(0.05,4)', \rho=0$; (d) $\bmu_1=(0,0)', \bmu_2=(0.05,4)',
\rho=0.9$. Figure \ref{contour} displays the empirical contour
plots from 1000 samples generated from the above four settings. In
all the settings, the underlying marginal distribution of $x_1$ is the same,
very close to a standard Gaussian. Thus it would be difficult to
disentangle the components based on a univariate analysis on $x_1$
alone. In contrast, in the presence of a small distance between
the means of $x_2$ in the two components, as in settings (a) and (b), there is already a mild
but noticeable improvement in the separation of the mixtures, reflected by the bend in the contour lines near the middle in Figures \ref{contour} (a) and (b). When the distance increases, as in settings (c) and (d), the separation of the components becomes
very visible. This is most striking in Figure \ref{contour}(d), where the two
components are completely separated. Given the same distance between the means of $x_2$ in the two components, higher conditional correlation within each component also appears to improve the disentanglement.

We first introduce some notations before presenting the main results that underlies the intuition. Given a sample of independent and identically distributed  (i.i.d.) random variables $\bx_1,...,\bx_n$ from the distribution \eqref{mixture}, we write the log likelihood as
\[
l(\btheta)=\sum_i^n \log f(\bx_i).
\]
The analytical forms of the score function and the Hessian matrix for arbitrary (finite) number of Gaussian mixtures with arbitrary dimension of observations are derived in \cite{Boldea09}. To simplify the analytical discussion, we focus on the simple case where the proportion $p$ and variance matrices $\bV_k$ are known, thus the only unknown parameters are $\btheta=\bmu=(\bmu_1,\bmu_2)'$.

Denote the score function by $\bq(\bmu)=\sum_i \bq_i(\bmu)$, where
\[
\bq_i(\bmu)=\frac{\partial \log f(\bx_i)}{\partial \bmu}=(\bq_{i1}, \bq_{i2}),
\]
and the Hessian matrix by $\bQ(\bmu)=\sum_i \bQ_i(\bmu)$, where
\[
\bQ_i(\bmu)=\frac{\partial^2 \log f(\bx_i)}{\partial \bmu \partial \bmu'}=
\left(
\begin{array}{cc}
\bQ_{i11} & \bQ_{i12} \\
\bQ_{i21} & \bQ_{i22}
\end{array}
\right).
\]
The maximum likelihood estimate (MLE) of the parameters, $\hat{\bmu}$, can be obtained via the EM algorithm by finding the solution to the system of equations of setting the score functions to 0. With the EM algorithm, the missing mixture membership is augmented  iteratively for each unit and the likelihood is maximized conditional on the augmentations. In likelihood-based approaches, the variance is usually estimated from the information matrix. If the model is correctly specified, the information matrix is defined by
\[
\mathcal{I}=-\bE(\bQ)=\bE(\bq \bq'),
\]
where the second equality holds because of the second-order regularity of $f$. The asymptotic variance of the MLE of $\bmu$ is $\mathcal{I}^{-1}$.

\subsection{Analytical results}
Our main analytical results are obtained through investigating the allocation probability $-$ the probability of unit $i$\ being in the group $k$:
\begin{equation}
\Pr(i\in k \mid \bx_i)= \frac{c f_{k}(\bx_{i})}{f(\bx_{i})}=\left\{ 1+\frac{\bar{c}~f_{\bar{k}}(\bx_{i})}{c~f_{k}(\bx_{i})}\right\} ^{-1}, \label{AR}
\end{equation}
where
\begin{equation*}
c=\left\{
\begin{array}{rl}
p,   &\text{for }k=1 \\
1-p, &\text{for }k=2%
\end{array}%
\right. , \quad
\bar{c}=\left\{
\begin{array}{rl}
1-p, &\text{for }  k=1 \\
p,   &\text{for }  k=2%
\end{array}%
\right. ,
\quad
\bar{k}=\left\{
\begin{array}{cl}
2, &\text{for }k=1 \\
1, &\text{for }k=2%
\end{array}%
\right. .
\end{equation*}
The allocation probability tends to 1 or 0 the better the mixture
disentanglement is, while it tends to $p$ or $1-p$ the worse the
mixture disentanglement is.

We first investigate the properties of the key term
$f_{\bar{k}}(\bx_{i})/f_{k}(\bx_{i})$ in \eqref{AR} for the
special case where the two components have the same variance
covariance matrix (homoscedasticity),  $\bV_1=\bV_2=\bV$. To
simplify discussion, we make the following transformations:
\begin{equation}
\bmu_{1}=\left(
\begin{array}{c}
0 \\
0%
\end{array}%
\right),
\quad
\bmu_{2}=\mathbf{d}=\left(
\begin{array}{c}
d_{1}=\mu_{12}-\mu_{11} \\
d_{2}=\mu_{22}-\mu_{21}%
\end{array}%
\right),
\quad
\bV=\left(
\begin{array}{cc}
\sigma _{1}^{2} & \rho ~\sigma _{1}\sigma _{2} \\
\rho ~\sigma _{1}\sigma _{2} & \sigma _{2}^{2}%
\end{array}
\right).
\label{parametric set}
\end{equation}%
Then the term $f_{\bar{k}}(\bx_{i})/f_{k}(\bx_{i})$ can be expressed as:
\[
f_{\bar{k}}(\bx_{i})/f_{k}(\bx_{i})=\exp \left\{ \mathbf{d}^{\prime }\mathbf{V}^{-1}(%
\bx_{i}-\bd/2)\right\}=\exp \left\{ h(\bx_i)\right\}.
\]
When $\bx_{i}$ belongs to component $k=1$, $h(\bx_{i})\sim
\normal(-\bd'\bV^{-1}\bd/2,~\bd'\bV^{-1}\bd)$, while when $\bx_{i}$ belongs to component $%
k=2$, $h(\bx_{i})\sim \normal(\bd'\bV^{-1}\bd/2,~\bd'\bV^{-1}\bd)$. Further, we can show that the probability of
$h(\bx_{i})<0$ when $\bx_{i}$ comes from group $k=1$, as well as the
probability of $h(\bx_{i})>0$ when $\bx_{i}$ comes from group $k=2$, increases with $\bd'\bV^{-1}\bd$ as
\begin{eqnarray*}
\Pr (h(\bx_{i})<0 \mid k=1)&=&\Phi \left(-\bE(h(\bx_{i}))\bVar(h(\bx_{i}))^{-1/2}\right)=\Phi ((\bd'\bV^{-1}\bd)^{1/2}/2),\\
\Pr (h(\bx_{i})>0 \mid k=2)&=&\Phi \left(\bE(h(\bx_{i}))\bVar(h(\bx_{i}))^{-1/2}\right)=\Phi ((\bd'\bV^{-1}\bd)^{1/2}/2).
\end{eqnarray*}
By the Bayes rule, we have
\begin{eqnarray*}
\Pr(k=1\mid h(\bx_{i})<0) &=& \frac{p~\Pr(h(\bx_{i})<0 \mid k=1)}{p~\Pr(h(\bx_{i})<0 \mid k=1) + (1-p)~\Pr(h(\bx_{i})<0 \mid k=2)} \nonumber\\
&=& \left[1+ \frac{1-p}{p} \left\{\frac{1}{\Phi ((\bd'\bV^{-1}\bd)^{1/2}/2)}-1 \right\}\right]^{-1}, \label{Bayes}
\end{eqnarray*}
and similar formula for $\Pr(k=2\mid h(\bx_{i})>0)$.

These results illustrate the critical role of $\bd'\bV^{-1}\bd$,
essentially a standardized distance between the two components, in
disentangling the components: As $\bd'\bV^{-1}\bd$ increases from
the minimum 0 to infinity, the probability $\Pr(k=1\mid
h(\bx_{i})<0)$ increases monotonically from its minimum $p$
towards the maximum 1, where the former is equivalent to a random
assignment of the group membership for each unit and the latter is
equivalent to the case when the group membership is known. Writing out
 $\bd'\bV^{-1}\bd$ using the parameters in \eqref{parametric set},
$$\bd'\bV^{-1}\bd=(d_{1}^{2}\sigma _{2}^{2}+d_{2}^{2}\sigma_{1}^{2}-2\rho\sigma_{1}\sigma_{1}d_{1}d_{2})/\{(1-\rho^2)\sigma _{1}^{2}\sigma _{2}^{2}\},$$
we have the following results.

\medskip
\noindent {\sc Result 1.}
\emph{For a bivariate mixture model of two Gaussian components with equal and known variance-covariance matrices, given the parameterization in equations \eqref{mixture} and \eqref{parametric set}}:
\begin{enumerate}
\item[(1)]\emph{For fixed values of $(d_1, \sigma _{1}, \sigma
_{2}, \rho)$, $\bd'\bV^{-1}\bd$ reaches its minimum at $d_2=\rho
d_1\sigma_2/\sigma_1$, and the minimum is $d_1^2/\sigma_1^2$,
which is the same value of $\bd'\bV^{-1}\bd$ in the univariate mixture model.}

\item[(2)] \emph{For fixed values of $(d_1, d_2, \sigma _{1},
\sigma _{2})$, $\bd'\bV^{-1}\bd$ reaches its minimum at two
mutually exclusive values of $\rho$: $d_2\sigma_{1}/d_1\sigma_2$
or $d_1\sigma_2/d_2\sigma_{1}$, and the minimum is either
$d_1^2/\sigma_1^2$(the same value of $\bd'\bV^{-1}\bd$ in the univariate mixture model) or
$d_2^2/\sigma_2^2$ (a value strictly greater than $d_1^2/\sigma_1^2$),
respectively.}

\item[(3)] \emph{For fixed values of $(d_1, \sigma _{1}, \sigma_{2}, \rho)$, the probability of
allocating unit $i$ to group $k$ when unit $i$ indeed belongs to
component $k$, $c f_{k}(\bx_{i})/f(\bx_{i})$, increases with
$d_{2}$ and \[\lim_{\left\vert d_{2}\right\vert \rightarrow \infty }\{c~f_{k}(\mathbf{x}_{i})/f(\mathbf{x}_{i})\mid ({i\in k})\}=1,\]
while for fixed values of $(d_1, d_2, \sigma _{1}, \sigma_{2})$, the probability of allocating unit $i$ to group $k$ when unit $i$ indeed belongs to component $k$, $c f_{k}(\bx_{i})/f(\bx_{i})$, increases with $\rho$ and
    \[\lim_{\left\vert \rho\right\vert \rightarrow 1 }\{c~f_{k}(\mathbf{x}_{i})/f(\mathbf{x}_{i})\mid ({i\in k})\}=1.\]}
\end{enumerate}

\noindent {\sc Proof.} See Appendix 1.

Result 1 states that, under correct model specification, the
standardized distance between the groups and thus the probability
of correctly assigning group membership for each unit from a
bivariate model is always greater than or equal to that from the
corresponding marginal univariate model, and it increases with the
distance between the group means of the second variable, and/or
the conditional correlation between the variables within
components.

The allocation probability is closely related to the information
matrix. Specifically, the score function
\begin{equation*}
\bq_{ik}(\bmu)=\partial \log f(\bx_{i})/\partial \bmu_{k}=c~\{f_{k}(\bx_{i})/f(\bx_{i})\}~\bV%
_{k}^{-1}(\bx_{i}-\bmu_{k}). \label{score}
\end{equation*}
Consider the consistent estimator for the information matrix $-$ the outer product of the scores evaluated at the MLE:
\begin{equation}
\mathcal{I}_{1}=\sum_{i}^{n}~\bq_{i}(\hat{\bmu})~\bq_{i}(\hat{\bmu})^{\prime }.
\end{equation}
The following result can be proved.

\medskip
\noindent {\sc Result 2.} \emph{For a bivariate mixture model of two Gaussian components with equal and known variance-covariance matrices (with parameterization in equations \eqref{mixture} and \eqref{parametric set}), given correct model specification, fixed $(d_1, \sigma _{1}, \sigma _{2}, \rho)$ and fixed sample size $n$,}
\begin{equation*}
\lim_{|d_2|\rightarrow \infty}\mathcal{I}_1=
\left(
\begin{array}{cc}
\mathbf{V}^{-1}\sum_{i\in k=1}~(\mathbf{x}_{i}-\hat{\bmu}_{1})(%
\mathbf{x}_{i}-\hat{\bmu}_{1})^{\prime }\mathbf{V}^{-1} & \mathbf{0}
\\
\mathbf{0} & \mathbf{V}^{-1}\sum_{i\in k=2}~(\mathbf{x}_{i}-\hat{\bmu}_{2})(\mathbf{x}_{i}-\hat{\bmu}_{2})^{\prime }\mathbf{V}^{-1}%
\end{array}
\right),
\end{equation*}
\emph{where $\hat{\bmu}_k$ is the MLE of $\bmu_k$, and the diagonal blocks are the outer products of the scores for $\bmu_k$ when the mixture membership for each unit is known. The same result holds when $|\rho|\rightarrow 1$, for fixed $(d_1, d_2, \sigma _{1}, \sigma _{2})$ and fixed sample size $n$.}

\noindent {\sc Proof.} See Appendix 2.

\medskip

Distinct from standard asymptotic results regarding increasing sample size, Result 2 is obtained with fixed $n$ but increasing values of $d_2$ or $\rho$. It implies that as the distance between the means of the secondary variable in two components or/and the conditional correlation between the two variables increases, the information number for the means of the primary variable converges to its maximum value $-$ the one from an analysis with the component labels known.

Intuitively, similar results also hold for mixtures with unequal variance-covariance matrices. However, general analytical results are difficult to obtain. We consider a second special case, where the two variables are conditionally independent in each group, that is, $\rho_1=\rho_2=0$, regardless of whether the variances $\sigma_1, \sigma_2$ are the same. Corresponding to Result 1 and 2, we have the following results.

\medskip
\noindent {\sc Result 3.}
\emph{For a bivariate mixture model of two Gaussian components with known variance-covariance matrices and $\rho_1=\rho_2=0$, given correct model specification, fixed values of $(d_1, \sigma _{11}, \sigma_{21}, \sigma_{12}, \sigma_{22})$ and fixed sample size $n$:}
\begin{enumerate}
\item[(1)] \emph{The probability of allocating unit $i$ to group $k$ when unit $i$ indeed belongs to component $k$, $c f_{k}(\bx_{i})/f(\bx_{i})$, increases with
$d_{2}$ and
\[\lim_{\left\vert d_{2}\right\vert \rightarrow \infty }\{c~f_{k}(\mathbf{x}_{i})/f(\mathbf{x}_{i})\mid ({i\in k})\}=1.\]}
\item[(2)] \emph{The estimated information matrix $\mI_1$}
\begin{equation*}
\lim_{|d_2|\rightarrow \infty}\mathcal{I}_1=
\left(
\begin{array}{cc}
\mathbf{V}_{1}^{-1}\sum_{i\in k=1}(\mathbf{x}_{i}-\hat{\bmu}_{1})(\mathbf{x}_{i}-\hat{\bmu}_{1})^{\prime}\mathbf{V}_{1}^{-1} & \mathbf{0}
\\
\mathbf{0} & \mathbf{V}_{2}^{-1}\sum_{i\in k=2}~(\mathbf{x}_{i}-\hat{\bmu}_{2})(\mathbf{x}_{i}-\hat{\bmu}_{2})^{\prime }\mathbf{V}_{2}^{-1}
\end{array}
\right),
\end{equation*}
\emph{where $\hat{\bmu}_{k}$ is the MLE of $\bmu_{k}$, and the diagonal blocks are the outer products of the scores for $\bmu_{k}$ when the mixture membership for each unit is known.}
\end{enumerate}

\noindent {\sc Proof.} See Appendix 3.

\medskip

These results are intuitive, because as the secondary outcome
distribution is increasingly separated between the two components
(increasing $d_2$), the component labels become clearer until
completely known, regardless of whether the primary outcome
distribution is well separated. A cross-dimensional comparison of
the information number with fixed $d_2$, on the other hand, may be
informative in practice, but is also more difficult to obtain.
Below we present a result derived with fixed $d_2$ for the special
case of equal $\bV$ and $\rho =0$.

\medskip
\noindent {\sc Result 4.} \emph{For a bivariate mixture model of
two Gaussian components with equal and known variance-covariance
matrices and $\rho=0$, given correct model specification and fixed
values of $(d_1, d_2, \sigma _{1}, \sigma _{2})$, the information
numbers for the means of the primary variable in $\mathcal{I}_1$
are larger than the corresponding ones from the univariate model for a large sample size $n$.}

\noindent {\sc Proof.} See Appendix 4.

\medskip

Result 4 is not a direct comparison of the estimated standard errors. Nevertheless, given $\rho =0$, results from simulations show the off-diagonal terms of $\mathcal{I}_1$ from the bivariate model quickly disappear with increasing $d_{2}$. Consequently the estimated standard errors for the means of the primary variable from the bivariate model can be approximated by the inverse of their information numbers, which can be easily shown to be lower than the estimated standard error from the corresponding marginal univariate model, given the positive definiteness of covariances matrices.

The above results are established assuming correct model specification. However, the information gain from utilizing secondary variables is obtained at the cost of having to specify more complex multivariate models. The number of parameters to be estimated in mixture models increases rapidly with the number of variables involved in the anaysis, increasing model uncertainty and also the possibility of misspecification. In particular,
multivariate normality is a much stronger assumption than univariate normality. It is therefore crucial to assess model assumptions in multivariate analysis. In the case of Gaussian mixtures, one way to assess normality and homoscedasticity is to apply the test of \cite{Hawkins81} to the clusters implied by the MLE. More discussions on this can be found in \cite{McLachlanBasford88}, Section 3.2, and \cite{McLachlan92}, Chapter 6.

Another benefit of introducing an auxiliary variable, which will be partly shown in the empirical analysis below, is that it tends to regularize the model. Mixture models with Gaussian (as well as most uni-modal distributions) components are not regular in the sense the ML regularity conditions for the likelihood function only hold locally, so that the likelihood function will generally have multiple roots, only one of which corresponds to the efficient likelihood estimator. The prevalence and the likelihood of spurious roots tends to disappear with the introduction of an auxiliary variable that it is highly associated with the mixture membership. For the means of the primary outcome of a mixture of two Gaussians, the gain is intuitive: upon inspecting the general formula of the observed information numbers calculated as the outer product of gradients. These are linear combinations of the squared scores and, for each component, the introduction of an auxiliary variable tends to annul the addends provided by the units belonging to the wrong component, and taking only those from the correct component. Discarding the wrong information results in an observed information matrix having the structure of a diagonal block matrix with null off-diagonal blocks and diagonal blocks equal to those of two regular Gaussian models. For a very entangled mixture of two Gaussian distributions the benefit in the standard error tends to be at most equivalent to multiply the sample size by the inverse of the mixture proportion, i.e., equivalent to doubling the sample size if the mixture proportion is equal to 0.5.

\section{Simulations}
General analytical results of cross-dimensional comparison for
mixture models with arbitrary number of components and dimensions
are difficult to establish; several analytical results were
however obtained for some special cases. We conduct now simulation
studies to investigate the small-sample behavior of bivariate
analysis and the corresponding marginal analysis of mixture models
under a wider range of settings. Specifically, we examine the
estimated standard error of the MLE for the component means of the
first variable $\mu _{1k}$ ($k=1,2$).

Besides $\mI_1$, we also consider the estimator of the variance
matrix based on the Hessian matrix of the likelihood:
\begin{equation}
\mathcal{I}_{2}= -\bQ(\hat{\bmu})= -\sum_i \bQ_i(\hat{\bmu}).
\end{equation}
If the model is correctly specified, the inverses $\mI_1^{-1}$ and $\mI_2^{-1}$ are both consistent estimators of the asymptotic variance of $\hat{\bmu}$. The closed-form Hessian matrix for $\bmu$ was derived in \cite{Boldea09}. Under model mis-specification, we will also consider the robust ``sandwich'' estimator \citep{Huber67}:
\begin{equation}
\mI_{3}^{-1}=\hat{\bVar}(\hat{\bmu})= \mI_2^{-1}\mI_1\mI_2^{-1},
\end{equation}
which is a consistent estimator for the variance, whether or not the model is correctly specified.

We consider three simulation settings, all with the sample size $n=500$ and the weight of component 1, $p=0.4$.
\begin{enumerate}
\item[S1.] \emph{Correctly specified model with known covariance matrices.} The data is generated from the bivariate Gaussian mixture density \eqref{mixture}-\eqref{input}, with $\bmu_1=(0,0)'$, $\bmu_2=(1, d_2)'$, and $\bV_1=\bV_2=\left(
\begin{array}{cc}
1 & \rho \\
\rho & 1
\end{array}
\right)$.
The parameters are estimated assuming the Gaussian mixture structure with $p$ and $\bV$ fixed at the true values.
\item[S2.] \emph{Correctly specified model with unknown covariance matrices}. The data is generated from the bivariate Gaussian mixture density \eqref{mixture}-\eqref{input} with the same true parameters as in S1. The parameters are estimated assuming the Gaussian mixture structure with unknown $p$ and $\bV$.
\item[S3.] \emph{Misspecified models (skewed with heavy tail) with unknown covariance matrices}. The data is generated from a mixture of two bivariate non-central \textit{t} distributions whose marginals have the same shape, using the following steps:
%\begin{enumerate}

 (i) Draw a sample of size $n=500$ from a Bernoulli distribution with $\Pr(i=1)=0.4$ and let $n_{1}$ be the number of time that $i=1$ and $n_{2}=n-n_{1}$.

 (ii) For $i=1,2$ draw $2n_{i}$ random number from the univariate
non-central $t$ distribution with degree of freedom $df=20$ and  non-centrality
parameter $\lambda =7$ using the formula: $t=(Z+\lambda)/(W/df)^{1/2}$, where $Z\sim \mbox{N}(0,1)$, $W\sim
\chi_{df}^{2}$.

 (iii) Standardize these random draws to have mean 0 and variance 1 (the
mean of the non-central \textit{t} with $df=20$ and $\lambda=7$ is: $\lambda ~\frac{df}{2}^{1/2}~\Gamma (\frac{df-1}{2})/\Gamma (%
\frac{df}{2})\approx 7.28$, and the variance is $\frac{df~(1+\lambda ^{2})}{df-2}-\frac{\lambda ^{2}~df}{2}\left\{ \Gamma (\frac{df-1}{2})/\Gamma (\frac{df}{2})\right\} ^{2}\approx 2.60$), and arrange the standardized numbers in bivariate vectors $\mathbf{\epsilon }_{i,1},...,\mathbf{\epsilon }_{i,n_{i}}$

 (iv) Transform $\mathbf{\epsilon }_{i,j }$ to $\mathbf{x}_{i,j }$ (for $j
=1,...,n_{i}$) by $\mathbf{x}_{i,j}=$ $\bmu_i+\mathbf{\epsilon }%
_{i,j }\bC_{i}$, with $ \bmu_i=(\bE(x_{1i}), \bE(x_{2i}))'$ and $\bC_{i}$ being the Choleski decomposition of the
desired correlation matrix $\bV=\bC_i\bC_i'$.
%\end{enumerate}

It is straightforward to show the above steps simulate the set $\left\{\mathbf{x}_{i,j}\right\}$ that satisfies $\bE(x_{11})=0, \bE(x_{12})=1, \bE(x_{21})=0$, $\rho(x_{1k},x_{2k})=\rho$ for $k=1,2$, and $\bVar(x_{mk})=1$ for $m,k=1,2$. %\comment{is this correct? how do you change $d_2$?}%

\end{enumerate}
For each setting, we conduct two series of simulations: (1) fixing
$\rho=0$ and  increasing the difference between the means of the
second variable $d_2$; (2) fixing $d_2=0$ and increase the
correlation between the two variables within each component
$\rho$. By Result 1.1, $d_2=0$ leads to the smallest allocation
probability when $\rho=0$, while by Result 1.2, $\rho=0$ leads to
the smallest allocation probability when $d_2=0$.

The MLEs of the parameters are obtained from the EM algorithm, and
the standard errors of the MLE are estimated from $\mI_{1}$,
$\mI_{2}$ and $\mI_3$. Following \cite{Boldea09}, for each setting
we obtain the Monte Carlo (MC) approximation to the true standard
error of $\hat{\mu}_{11}$ (and $\hat{\mu}_{12}$) as the standard
deviations of the empirical distributions of the MLE
$\hat{\mu}_{11}$ and $\hat{\mu}_{12}$ from $R=10000$ replicates,
each of sample size $n=500$. The subsequent estimated standard
errors, from $\mI_{1}$, $\mI_{2}$, other than $\mI_3$ for S3, are
assessed in terms of bias and root mean square error (RMSE) to the
``true standard errors'', calculated from 1000 replicates.

The estimated standard errors of the MLE for $\hat{{\mu}}_{11}$
from estimating a bivariate normal mixture models versus those
from estimating the marginal univariate model are summarized in
Table \ref{tab:sim1}, \ref{tab:sim2}, \ref{tab:sim3} under
settings S1, S2, S3, respectively. The last column of each table
reports the estimated allocation rate, which is an estimate of the
proportion of units that are correctly allocated to the
components. The allocation rate is a useful indicator for
quantifying mixture disentanglement; it is here estimated by
averaging the higher probability of unit $i$\ being in the group
$k$\ calculated at the MLE \citep{McLachlanBasford88}:
$\left\{\sum_{i}\max_{k}\,\Pr (i\in k~|~\mathbf{x}_{i})\right\}
/n$. The lower bound for the estimated allocation rate is $0.5$;
low values correspond to poor mixture disentanglements, and vice
versa.

\begin{table}[!h]
\begin{center}
\begin{footnotesize}
\begin{tabular}{ccccccccccccc}
\\ \hline\hline
&  & \multicolumn{4}{c}{$\mathcal{I}_{1}$} && \multicolumn{4}{c}{$\mathcal{I}_{2}$} && \\
\cline{3-6} \cline{8-11}
&$d _{2}$ or $\rho$    & $|\mbox{bias}|$ & mean $\widehat{\mbox{se}}$ & RMSE& (*)    && $|\mbox{bias}|$    & mean $\widehat{\mbox{se}}$ & RMSE & (*) && $\widehat{\mbox{AR}}$  \\ \hline
univ    &-   & $1.8 e^{-3}$ & $1.1 e^{-1}$ &$1.5 e^{-2}$ & -      && $2.5 e^{-3}$ &$1.1 e^{-1}$  &$1.3 e^{-2}$  &-      && $.703$\\
\hline
          &$0$ & $3.4 e^{-3}$ & $1.1 e^{-1}$ &$1.3 e^{-2}$ & $345$ && $3.2 e^{-3}$ & $1.1 e^{-1}$ & $1.5 e^{-2}$ & $531$ && $.694$ \\
biv       &$1$ & $1.1 e^{-3}$ & $1.0 e^{-1}$ &$6.2 e^{-3}$ & $993$ && $1.6 e^{-3}$ &$1.0 e^{-1}$  &$5.7 e^{-3}$  & $911$ && $.769$ \\
($\rho=0$)&$3$ & $1.9 e^{-4}$ & $7.7 e^{-2}$ &$4.3 e^{-3}$ & $998$ && $8.9 e^{-4}$ &$7.6 e^{-2}$  &$2.4 e^{-3}$  & $1000$&& $.945$ \\
          &$5$ & $1.2 e^{-4}$ & $7.2 e^{-2}$ &$4.0 e^{-3}$ & $1000$&& $7.2 e^{-4}$ &$7.2 e^{-2}$  &$2.0 e^{-3}$  & $1000$&& $.996$ \\
          &$50$& $6.5 e^{-4}$ & $7.1 e^{-2}$ &$4.0 e^{-3}$ & $999$ && $6.6 e^{-5}$ &$7.1 e^{-2}$  &$1.8 e^{-3}$  & $1000$&& $1.00$\\
\hline
            &$.50$ & $9.6 e^{-4}$ & $1.1 e^{-1}$ & $9.7 e^{-3}$ & $839$ && $1.8 e^{-3}$ & $1.1 e^{-1}$ & $9.8 e^{-3}$ & $740$ && $.726$\\
biv         &$.75$ & $1.6 e^{-3}$ & $9.8 e^{-2}$ & $5.9 e^{-3}$ & $999$ && $8.6 e^{-4}$ & $9.8 e^{-2}$ & $4.7 e^{-3}$ & $948$ && $.783$\\
($d _{2}=0$)&$.90$ & $4.4 e^{-4}$ & $8.5 e^{-2}$ & $4.7 e^{-3}$ & $1000$&& $2.5 e^{-4}$ & $8.4 e^{-2}$ & $2.6 e^{-3}$ & $1000$&& $.878$ \\
            &$.99$ & $6.3 e^{-4}$ & $7.1 e^{-2}$ & $4.3 e^{-3}$ & $999$ && $6.0 e^{-5}$ & $7.1 e^{-2}$ & $1.9 e^{-3}$ & $1000$&& $1.00$\\
\hline
\hline
\end{tabular}
\end{footnotesize}
\caption{The estimated standard error of the MLE of $\mu_{11}$ from bivariate analysis and corresponding marginal analysis under the simulation setting S1. The upper panel is with fixed $\rho=0$ and varying $d_2$, and the lower panel is with fixed $d_2=0$ and varying $\rho$. Here ``$|\mbox{bias}|$" stands for absolute bias; ``mean $\widehat{\mbox{se}}$" stands for mean of the estimated s.e. of $\hat{\mu}_{11}$, (*) is number of times that, over the 1000 replications, the bivariate estimated s.e. of $\hat{\mu}_{11}$ is smaller than the univariate one, and ``$\widehat{\mbox{AR}}$'' is the estimated allocation rate.} \label{tab:sim1}
\end{center}
\end{table}

\begin{table}[!h]
\begin{center}
\begin{footnotesize}
\begin{tabular}{ccccccccccccc}
\\ \hline\hline
&  & \multicolumn{4}{c}{$\mathcal{I}_{1}$} && \multicolumn{4}{c}{$\mathcal{I}_{2}$} &&\\
\cline{3-6} \cline{8-11}
&$d _{2}$ or $\rho$ &$|\mbox{bias}|$ & mean $\widehat{\mbox{se}}$ & RMSE & (*) && $|\mbox{bias}|$ & mean $\widehat{\mbox{se}}$ & RMSE & (*) && $\widehat{\mbox{AR}}$  \\ \hline
univ    &-   & $1.7 e^{-3}$ & $1.1 e^{-1}$ &$1.5 e^{-2}$ & -      && $2.5 e^{-3}$ &$1.1 e^{-1}$  &$1.3 e^{-2}$  &-      && $.893$\\
\hline
            &$0$ & $3.4 e^{-1}$ & $ 2.8e^{-1}$ &$ 4.1e^{-1}$ & $621$  && $ 3.4e^{-1}$ &$ 2.8e^{-1}$  &$ 6.3e^{-1}$  & $573$ && $.862 $ \\
biv         &$1$ & $2.8 e^{-1}$ & $ 3.0e^{-1}$ &$ 4.0e^{-1}$ & $587$  && $ 3.2e^{-1}$ &$ 2.5e^{-1}$  &$ 4.0e^{-1}$  & $557$ && $.871 $ \\
($\rho=0$)  &$3$ & $2.1 e^{-3}$ & $ 8.8e^{-2}$ &$ 9.7e^{-3}$ & $804$  && $ 2.8e^{-3}$ &$ 9.3e^{-2}$  &$ 4.8e^{-2}$  & $782$ && $.946 $ \\
            &$5$ & $2.4 e^{-4}$ & $ 7.2e^{-2}$ &$ 4.6e^{-3}$ & $839$  && $ 3.0e^{-3}$ &$ 7.5e^{-2}$  &$ 4.0e^{-2}$  & $819$ && $.995 $ \\
            &$50$& $6.4 e^{-4}$ & $ 7.2e^{-2}$ &$ 4.3e^{-3}$ & $823$  && $ 1.1e^{-3}$ &$ 7.2e^{-2}$  &$ 1.6e^{-2}$  & $807$ && $1.00$\\
\hline
            &$.50$ & $3.1 e^{-1}$ & $3.0e^{-1}$ &$ 4.1e^{-1}$ & $606$  && $ 3.4e^{-1}$ &$ 2.7e^{-1}$  &$ 4.8e^{-1}$  & $558$&& $.863 $  \\
biv         &$.75$ & $2.4 e^{-1}$ & $2.9e^{-1}$ &$ 4.5e^{-1}$ & $573$  && $ 2.7e^{-1}$ &$ 2.6e^{-1}$  &$ 3.5e^{-1}$  & $533$&& $.867 $  \\
($d _{2}=0$)&$.90$ & $4.9 e^{-2}$ & $1.6e^{-1}$ &$ 7.1e^{-2}$ & $670$  && $ 4.2e^{-2}$ &$ 1.7e^{-1}$  &$ 1.1e^{-1}$  & $630$&& $.887 $  \\
            &$.99$ & $6.5 e^{-4}$ & $7.2e^{-2}$ &$ 4.4e^{-3}$ & $846$  && $ 2.6e^{-3}$ &$ 7.3e^{-2}$  &$ 6.2e^{-2}$  & $839$&& $1.00$  \\
\hline
\hline
\end{tabular}
\end{footnotesize}
\caption{The estimated standard error of the MLE of $\mu_{11}$ from bivariate analysis and corresponding marginal analysis under the simulation setting S2. The upper panel is with fixed $\rho=0$ and varying $d_2$, and the lower panel is with fixed $d_2=0$ and varying $\rho$.} \label{tab:sim2}
\end{center}
\end{table}

%\begin{sidewaystable}[!h]
\begin{table}[!h]
\begin{center}
\begin{footnotesize}
\begin{tabular}{ccccccccccccccc}
\\ \hline\hline
&  & \multicolumn{3}{c}{$\mathcal{I}_{1}$} && \multicolumn{3}{c}{$\mathcal{I}_{2}$} && \multicolumn{3}{c}{$\mathcal{I}_{3}$} &&
\\ \cline{3-5} \cline{7-9} \cline{11-13}
& $d _{2}$ or $\rho$ & $|\mbox{bias}|$ & mean $\widehat{\mbox{se}}$ & RMSE && $|\mbox{bias}|$ & mean $\widehat{\mbox{se}}$ & RMSE  && $|\mbox{bias}|$ & mean $\widehat{\mbox{se}}$  & RMSE  && $\widehat{\mbox{AR}}$\\ \hline
       &$4$ & $4.2 e^{-3}$ & $8.1e^{-2}$ &$ 9.5e^{-3}$   && $ 9.6e^{-4}$ &$ 7.6e^{-2}$  &$ 1.9e^{-2}$  && $ 1.4e^{-2}$ &$ 9.0e^{-2}$  &$ 6.6e^{-2}$ && $.986$   \\
$\rho=0$&$5$&$7.0 e^{-3}$ & $7.9e^{-2}$ &$ 1.1e^{-2}$ && $ 2.1e^{-3}$ &$ 7.4e^{-2}$  &$ 1.9e^{-2}$  && $ 1.7e^{-2}$ &$ 8.9e^{-2}$  &$ 7.3e^{-2}$ && $.996$  \\
    &$50$& $7.3 e^{-3}$ & $7.8e^{-2}$ &$ 1.1e^{-2}$   && $ 2.1e^{-4}$ &$ 7.1e^{-2}$  &$ 9.8e^{-3}$  && $ 2.5e^{-3}$ &$ 7.4e^{-2}$  &$ 2.3e^{-2}$ && $1.00$\\
\hline
        &$0.50$ & $2.5 e^{-3}$ & $8.0e^{-2}$ &$ 9.0e^{-3}$   && $ 4.4e^{-3}$ &$ 7.3e^{-2}$  &$ 1.8e^{-2}$  && $ 1.7e^{-3}$ &$ 8.0e^{-2}$  &$ 4.7e^{-2}$ && $.987$   \\
$d_2=4$ &$0.75$ & $6.5 e^{-3}$ & $7.9e^{-2}$ &$ 1.1e^{-2}$   && $ 9.6e^{-4}$ &$ 7.2e^{-2}$  &$ 2.0e^{-2}$  && $ 6.0e^{-3}$ &$ 7.9e^{-2}$  &$ 6.0e^{-2}$ && $.995$   \\
 &$0.90$ & $7.1 e^{-3}$ & $7.8e^{-2}$ &$ 1.0e^{-2}$   && $ 1.5e^{-3}$ &$ 7.3e^{-2}$  &$ 2.1e^{-2}$  && $ 1.0e^{-2}$ &$ 8.1e^{-2}$  &$ 6.9e^{-2}$ && $.999$ \\
 &$0.99$& $6.6 e^{-3}$ & $7.8e^{-2}$ &$ 9.9e^{-3}$   && $ 1.3e^{-3}$ &$ 7.1e^{-2}$  &$ 1.6e^{-2}$  && $ 1.7e^{-3}$ &$ 7.3e^{-2}$  &$ 4.7e^{-2}$ && $1.00$\\
\hline
\hline
\end{tabular}
\end{footnotesize}
\caption{The estimated standard error of the MLE of $\mu_{11}$ from bivariate analysis, under the simulation setting S3. The upper panel is with fixed $\rho=0$ and varying $d_2$, and the lower panel is with fixed $d_2=4$ and varying $\rho$.} \label{tab:sim3}
\end{center}
%\end{sidewaystable}
\end{table}

When the model is correctly specified with known variance (setting
S1), as predicted by the analytical results, the bivariate
analysis nearly always outperforms the univariate analysis, and
the improvement increases as the distance between the two mixture
components of the secondary variable or the correlation between
the two variables increases. Here the estimator $\mI_2$ leads to
comparable standard errors but smaller bias (thus smaller MSE)
than $\mI_1$. Interestingly, the ratio between the bivariate and the univariate mean $\widehat{\mbox{se}}$ has a lower bound of size $7.1 e^{-2}/1.1 e^{-1}\cong\sqrt{0.4}$ i.e. $\frac{\sqrt{n}}{\sqrt{n/p}}=\sqrt{p}$. Thus, the reduction of the s.e. is equivalent to a reduction due to increase the sample size by the inverse of the mixture proportion. Results for the estimated s.e. of the MLE of $\hat{\mu}_{12}$ and for alternative sample sizes, not reported here, confirm this evidence.

When the model is correctly specified but with unknown variance
(setting S2), the bivariate analysis still leads to smaller
standard errors than the univariate analysis in at least 60\% of
the time, and this rate increases to over 80\% as $d_2$ increases
to 3. But unlike in setting S1, here $\mI_2$ leads to comparable
standard error but larger bias (thus larger MSE) than $\mI_1$.
Interestingly, in both settings S1 and S2, the improvement in bias
and MSE of bivariate analysis appears to plateau after $d_2$
reaches 5 despite the estimated allocation rate continuing to
increase with $d_2$. This illustrates that, in practice, a
secondary variable with even modest distance between the two
components is sufficent to provide noticeable improvement.

For setting S3, the ``sandwich'' estimator $\mI_{3}$ yields standard errors for the MLE that are robust to specification error. However, it ignores bias, which may be appreciable, so that results can be misleading \citep[e.g.][]{Freedman06}.
Consequently, we have considered only
values of $d_2$ and $\rho$ such that the pseudo-MLE provides a
good approximation to the true data model: the assessments have
been carried out when the bias $\bar{\hat{\mu}}_{11}-\bE(X_{11})$
is low, namely less than 0.03 (with $\bar{\hat{\mu}}_{11}$ be the
average of the empirical distribution of the MLE of
$\hat{\mu}_{11}$ over the $10000$ replicates). As a result, we
do not assess the performance of bivariate versus univariate
estimators this time, given the bad pseudo-MLE approximation
obtained for the latter (the bias over the 10000 replicates is
$\bar{\hat{\mu}}_{11}-\bE(X_{11})=0.305$). Moreover, the analysis
for increasing values of $\rho$ has been carried out by fixing the
distance $d_2$ to 4 instead of 0, since the posing of $d_2$ to 0
would have resulted in a good approximation of the pseudo-MLE only
when $\rho=0.999$. Table \ref{tab:sim3} shows the outer product
estimator $\mI_1$ leads to smaller standard error and comparable
bias, thus a better performance in term of RMSE, than both $\mI_2$
and the sandwich estimator $\mI_3$. The extra advantage of the bivariate analysis when the underlying model is incorrect is in the great reduction of bias, compared to the univariate case, leading to a really robust inference (more pronounced for increasing $d_2$ than for increasing $\rho$). The large value of the bias obtained for the univariate analysis shows it fails to provide a good approximation to the true data model, and signals an analysis of resulting MLE of $\mu_{11}$ would be misleading.

\section{Real applications}
\subsection{Crab data}
The crab data of the genus \emph{Leptograpsus variegatus},
originally collected by \cite{Campbell74}, has been often analyzed
in the literature of multivariate mixture models
\citep[e.g.][]{Ripley96, McLachlanPeel98,McLachlanPeel00}. Here we
focus the sample of $n=100$ blue crabs, with $n_1=50$ males and
$n_2=50$ females, corresponding to the two components with
component labels known. Each specimen has measurements (in mm) on
the width of the front lip (FL), the rear width (RW), the length
along the midline (CL), the body depth (BD), and the carapace
width (CW). We use the data to conduct cross-dimensional
comparison of the mixture models with more than two variables.
While Hawkins' test suggests both normality and homoscedasticity
assumptions to be reasonable here, \cite{McLachlanPeel00} found
that homoscedasticity may lead to inferior model fitting. For
illustration purpose, we consider the hypothetical setting that RW
is of primary interest and all other variables are secondary. We
performed three clustering analyses, ignoring the known component
labels: In the first, we fitted a univariate mixture model to RW,
in  the second we fitted a bivariate model to RW and CL, and in
the third we fitted  trivariate models to RW and CL with either
FL, or BD, or CW as an additional third variable; all the models
were with two Gaussian components and heterogeneous covariance
matrices. The MLEs of the parameters were obtained running the EM
algorithm with several random starting values. The labelling of
mixtures components are those obtained by setting the starting
values in both analyses as the component-specific sample means
(e.g., 11.72 and 12.14 for RW; 32.01 and 28.10 for CL), variances
(e.g., 4.46 and 5.95 for RW; 53.42 and 35.04 for CL) and
covariances. The results are reported in Table \ref{tab:crab_RW}.
In the univariate analysis, the EM converged to a spurious maximum
point, with $\hat{\sigma}_{\mbox{{\tiny RWmales}}}^{2}=.68$,
resulting from a group of eight outliers being erroneously
identified as a component. As a consequence, the A.R. was very low
(.10 for the males and .50 overall). The bivariate analysis reduced the adverse effect
of outliers, leading to $\hat{\sigma}_{\mbox{{\tiny
RWmales}}}=2.82$ and the overall A.R. improves from .50 to .87.
Also the standard errors from all three estimators improved for
the males (the comparison for females is not meaningful due to the
spurious point). The significant improvement is as expected from
the theoretical results because the empirical correlation (given
the true labels) between RW and CL is very high for both males
(.977) and females (.987). The trivariate analyses lead to
comparable results as the bivariate one. This plateau in
performance is not surprising since the information gain from
adding variables is obtained at the price of the extra uncertainty
in estimating more parameters; the latter can outweigh the former
especially when a lower-dimensional analysis already produces
accurate results.

\begin{table}[!h]
\begin{center}
\begin{tabular}{llccccccccc}
\hline\hline
&& \multicolumn{2}{c}{MLE} &&\multicolumn{3}{c}{s.e. for $\hat{\mu}_{\mbox{{\tiny RW}}}$} && \multicolumn{2}{c}{A.R.} \\ \cline{3-4} \cline{6-8} \cline{10-11}
&& $\hat{\mu}_{\mbox{{\tiny RW}}}$ & $\hat{\sigma}_{\mbox{{\tiny RW}}}^{2}$ && $\mathcal{I}_{1}$ & $\mathcal{I}_{2}$ & $\mathcal{I}_{3}$ && component &overall  \\ \hline
                     &univ          & 7.97 &0.68 && .83 & .65 & .53 && .100& .500  \\
                     &biv  (CL)     &12.40 &2.82 && .36 & .38 & .35 && .740& .870  \\
{male}               &triv (CL,FL)  &12.43 &2.72 && .34 & .32 & .32 && .740& .870 \\
                     &triv (CL,CW)  &12.44 &2.73 && .43 & .32 & .39 && .720& .860 \\
                     &triv (CL,BD)  &12.32 &2.81 && .38 & .33 & .35 && .800& .900 \\
\hline
                        &univ        &12.27 &4.04  && .36 & .31 & .29 && .900&.500  \\
                        &biv (CL)    &11.63 &6.40  && .34 & .34 & .34 && 1.00&.870  \\
{female}                &triv (CL,FL)&11.61 &6.41  && .39 & .33 & .34 && 1.00&.870 \\
                        &triv (CL,CW)&11.61 &6.39  && .36 & .33 & .37 && 1.00&.860 \\
                        &triv (CL,BD)&11.66 &6.58  && .39 & .35 & .34 && 1.00&.900 \\
\hline\hline
\end{tabular}
\caption{The estimated mean and variance of RW, and the standard error of the mean for male and female crabs. In the bivariate model, CL is as the second variable; in the trivariate analyses, FL, CW, BD are separately used as the third variable besides RW and CL. The sample mean of RW for the males and females is 11.72 and 12.14, respectively, and the sample variance of RW for the males and females is 4.46 and 5.95, respectively.} \label{tab:crab_RW}
\end{center}
\end{table}

Besides RW, we have also run similar analysis with CL as the primary outcome. No spurious point was detected. The standard errors of the cluster means estimated from all the three estimators reduced significantly (60\% in males and 80\% in females) from the univariate to the bivariate analysis and the overall A.R. increased from .60 to .87. Same as before, the trivariate analyses did not provide further improvement (details are omitted here).

In the presence of multiple candidate secondary variables, we suggest the
following selection procedure: first, conduct a normality test
(e.g., Hawkins') for each bivariate pair of the primary variable
and one secondary variable and select the ones deemed normal;
second, for each selected pair perform a bivariate mixture
analysis and, given the estimated labels, calculate the empirical
within-component correlations and the distance between the
component means of the secondary variable; third, choose the
secondary variable that gives the highest absolution correlations
or/and distances.

\subsection{Educational cost of World War II}
The second application arises from the Instrumental Variable (IV) approach in causal inference \citep{Angrist96}, which inherently defines a mixture structure as shown later. An instrumental variable or instrument is a variable that is correlated with the treatment variable, but does not have a direct effect on the outcome, only indirectly through the treatment variable. The instrumental variable is often viewed as defining a natural experiment. \cite{IchinoWinterEbmer04} used the IV approach to evaluate the long-run educational effect of World War II on earnings. In particular, they used the cohort of birth as an instrument ($Z$): $Z=1$ for individuals born between 1930 and 1939 (these individuals were in primary school age during the war and the immediately following period) and $Z=0$ otherwise. It is reasonable to assume that which year an individual was born is random (by nature) and does not directly affect one's earnings later in life once accounting for the secular trend towards higher earnings, but it can indeed affect the education level an individual received ($D=1$ for poorly educated, $D=0$ otherwise) due to the intervention of war, which in turn affects the earnings later. The population can be divided into four latent subpopulations according to an individual's potential (counterfactual) educational levels under different values of the instrument:
\begin{enumerate}
\setlength{\itemsep}{0pt}
\item \emph{Always-poorly educated} ($G=a$): individuals who would obtain low education levels irrespective of the cohort of birth;
\item \emph{Never-poorly educated} ($G=n$): individuals who would obtain high education levels irrespective of the cohort of birth;
\item \emph{Compliers} ($G=c$): individuals who would obtain low education level if born in the decade immediately before war, but would obtain high education level if not born in the decade immediately before war;
\item \emph{Defiers} ($G=d$), individuals who would obtain high education level if born in the decade immediately before war but low education level otherwise.
\end{enumerate}
It is usually reasonable to rule out defiers in practice. And the estimand of interest lies in the effect of war on earnings for compliers, known as the compliers average causal effect (CACE). %\citep{Angrist96}.

%Therefore, compliers are the only group comprised of individuals whose educational levels depend on the cohort of birth, and consequently the target of the analysis is to estimate

%To address the unobserved confounding due to self-selection of individuals to level of education, the authors adopted the Instrumental Variable (IV) approach, which will be shown to be an inherently mixture problem.
%An important long-run cost of war is the disruption of educational processes. Bombings, fighting, disruptions and related consequences made it difficult to physically access schools, therefore to achieve the desired level of education for part of the early school age population. Under this perspective,

The inferential challenge is that the individual's subclass membership is not always observed. Specifically, given the structural assumptions in \cite{Angrist96}, individuals for which we observe $(Z=0,D=1)$ are never-poorly educated, and those for which $(Z=0,D=1)$ are always-poorly educated. But individuals with $(Z=1,D=1)$ or $(Z=0,D=0)$ consists of two different mixtures, that are a mixture of always-poorly educated and compliers, and a mixture of never-poorly educated and compliers respectively, as shown in Table \ref{tab:compl_status}. For continuous outcomes, \cite{Imbens97} proposed the following mixture model for the above formulation of instrumental variable:
\begin{eqnarray}
f(x,d,z)&=& \one_{\{D=1,Z=0\}}\cdot (1-\pi )\cdot \omega _{a}\cdot \phi_{a0}(x)+\one_{\{D=0,Z=1\}}\cdot \pi \cdot \omega _{n}\cdot \phi_{n1}(x) \nonumber\\
&&+ \one_{\{D=1,Z=1\}}\cdot \pi \cdot \{\omega _{a}\cdot \phi_{a1}(x)+\omega _{c}\cdot \phi_{c1}(x)\} \nonumber\\
&&+\one_{\{D=0,Z=0\}}\cdot (1-\pi )\cdot \{\omega _{n}\cdot \phi_{n0}(x)+\omega
_{c}\cdot \phi_{c0}(x)\}, \label{WWII}
\end{eqnarray}
where $f(x,d,z)$ is the density of the outcome $x$ given the observed instrument $z$ and education level $d$, $\one_{\{A\}}$ is an indicator function of set $A$, $\pi$ is the probability $P(Z=1)$, $\omega _{g}$ is the mixing probability, which is the probability of an individual being in the $g$ group $P(G=g)$ for $g=a, c, n$, and  $\phi_{gz}(x)=\phi_{gz}(x;\mu_{gz},\sigma_{gz})$ is the (normal) outcome distribution for a unit in the $g$ group that is
assigned to the treatment $z$. The mixture structure is clearly shown in two last factors in (\ref{WWII}), which are linked to each other by means of the two parameters $\pi$ and $\omega_{c}$, and therefore two separate analyses of the mixtures would lead to different results compared to the joint analysis of (\ref{WWII}).

\begin{table}[!h]
\begin{center}
\begin{tabular}{ccc}
\hline\hline
& $Z=0$ & $Z=1$ \\ \hline
$D=0$ & never-poorly educated and compliers & never-poorly educated \\
$D=1$ & always-poorly educated & always-poorly educated and compliers \\
\hline\hline
\end{tabular}
\caption{Composition of the population classified by $D$ and $Z$.} \label{tab:compl_status}
\end{center}
\end{table}

Our analysis uses the same data of \cite{IchinoWinterEbmer04}, collected from the wave 1986 of the German Socio-Economic Panel from which we consider only males born between 1925 and 1949. They defined the long-run educational cost of World War II on earnings in Germany as the average earnings loss experienced by those individuals who received less education because they were about in primary school age during the war or the immediately following years.  To account for increasing trend of earnings with respect to age, following these authors, the primary outcome $X$ is defined as the residual of a regression of natural log of average hourly earnings observed in 1986 in Germany on a cubic polynomial in age. The log transformation for income is usually adopted in the labor economics in order to induce normality in such otherwise asymmetric variable. To account for decreasing trend of earnings with respect to age, the ``treatment'' $D$ is defined to be equal to one if the individual's residual of a regression of years of education on a cubic polynomial in age is smaller than the residuals' sample average (poorly educated) and zero otherwise. The instrument $Z$ was defined earlier. In addition, we select an auxiliary variable: the hours worked per week. We will compare the results from fitting the univariate version of model (\ref{WWII}) to earnings, versus those obtained from fitting the bivariate version of model (\ref{WWII}) to earnings and the auxiliary variable.

We first eliminate the multivariate outliers detected on the initial sample of 1163 units. A subsequent visual checking of the histograms of the auxiliary variable for the two non-mixtures factors in (\ref{WWII}) reveals a deviation from normality due to a slight bimodality. Individuals presenting very high level of hours worked have been consequently eliminated so that the final sample size results in 993 units. As shown in \cite{Mercatanti13}, despite model (\ref{WWII}) is identified, the main problem associated with a likelihood analysis arises from the possibility of having multiple roots for the likelihood equations, which results from the two mixtures of distributions being involved. We adopt here the proposed solution to identify the Efficient Likelihood Estimator (ELE) for parameters in (\ref{WWII}), under heteroscedastic conditions for the mixtures, as the local maximum likelihood point closest to the method of moments estimate of the mixing probabilities.

Table \ref{tab:WWII} reports the results of the univariate versus the bivariate analysis. The parameters related to the groups of units for which the subclass membership is observed (units with $D=1$ and $Z=0$, or $D=0$ and $Z=1$) do not involve mixture structure, and unsurprisingly their estimated values ($\hat{\mu}_{a0}$, $\hat{\sigma}_{a0}$, $\hat{\mu}_{n1}$, $\hat{\sigma}_{n1}$) and standard errors do not change from the univariate to the bivariate analysis. Significant reductions in the estimated standard errors have been obtained for the parameters of the mixture of always-poorly educated and compliers: this is a very entangled mixture for which the contribution of the secondary outcome is decisive to sharpen the inference. In particular for the group of compliers, for which the estimated standard errors show a reduction of about 55\% in $\hat{\mu}_{c1}$, and 60\% in $\hat{\sigma}_{c1}$. Moreover the most important estimand in this study, that is the average causal effect on earnings for compliers, $\hat{\mu}_{c1}-\hat{\mu}_{c0}$ (individuals whose educational choices were affected by the war), shows a reduction in its standard error ranging between 24\%, by $\mathcal{I}_{3}$, to 32\%, by $\mathcal{I}_{2}$, due essentially to the strong reduction observed for $\hat{\mu}_{c1}$. This leads to a decrease in the \textit{p}-value for this quantity from about $0.700$ for the univariate case to $0.054$ ($\mathcal{I}_{1}$), $0.048$ ($\mathcal{I}_{2}$), $0.081$ ($\mathcal{I}_{3}$) for the bivariate case. The estimated standard errors for the rest of the parametric set show lighter reductions apart from the slight increases in $\hat{\sigma}_{c0}$ and $\hat{\mu}_{a1}$.

Interestingly, another advantage of the bivariate analysis in this example emerges from the analysis of the local maximum likelihood points detected. The ELEs reported in Table \ref{tab:WWII} correspond to the roots closest to the method of moments estimates of the mixing probabilities. The second closest root detected for the univariate case reports parameters values similar of that obtained for the ELE in the bivariate case (even if with generally larger estimated standard errors). The mixture composed by always-poorly educated and compliers is very entangled in the univariate case; this complicates the analysis so that this solution remains confused with others local roots. The more effective disentanglement of the mixture allowed by the introduction of the secondary outcome succeeds in highlighting this solution as the ELE.

The practical interpretation of the results has to account for the definition of the outcome as log of earnings. This means the estimated effects are not differences in average amounts of money, but they are semi-elasticities, i.e. they show the approximate average percentage changes in earnings between groups of individuals classified by the cohort of birth. As expected the estimated effect for compliers is negative: the earnings are on average 25.73\% lower for compliers who were affected by war because in primary school age during that period. The effect for always-poorly educated is substantially zero, while that for never-poorly educated results is positive -- this is not surprising because it is reasonable to think that never-poorly educated individuals born between 1930 and 1939 took advantage of the lower average education level in their cohort by experiencing less competitive labour market conditions during their adulthood, thus increasing their average earnings.

\begin{table}[!h]
\begin{center}
\begin{tabular}{lccccccccc}
\hline\hline
& \multicolumn{4}{c}{Univariate case} &  & \multicolumn{4}{c}{Bivariate case}
\\ \cline{2-5}\cline{7-10}
& ELE & \multicolumn{3}{c}{s.e.} &  & ELE & \multicolumn{3}{c}{s.e.} \\   \cline{3-5} \cline{8-10}
&  & $\mathcal{I}_{1}$ & $\mathcal{I}_{2}$ & $\mathcal{I}_{3}$ &  &  & $\mathcal{I}_{1}$ & $\mathcal{I}_{2}$ & $\mathcal{I}_{3}$ \\ \hline
$\hat{\omega}_{a}$ & 0.7316 & 0.0293 & 0.0293 & 0.0293 &  & 0.7301 & 0.0293 & 0.0288 & 0.0290  \\
$\hat{\omega}_{n}$ & 0.2075 & 0.0192 & 0.0193 & 0.0196 &  & 0.2099 & 0.0186 & 0.0180 & 0.0186\\
$\hat{\omega}_{c}$ & 0.0608 & 0.0239 & 0.0239 & 0.0246 &  & 0.0599 & 0.0230 & 0.0206 & 0.0220 \\
$\hat{\mu}_{a0}$  & -0.1229 & 0.0142 & 0.0142 & 0.0143 &  & -0.1229& 0.0142 & 0.0139 & 0.0143 \\
$\hat{\mu}_{a1}$ & -0.1333 & 0.0171 & 0.0174 & 0.0180 &  & -0.1202 & 0.0185 & 0.0174 & 0.0194\\
$\hat{\mu}_{a1}-\hat{\mu}_{a0}$ & -0.0104 & 0.0222 & 0.0228 & 0.0234&  & 0.0027 & 0.0233 & 0.0225 & 0.0257 \\
$\hat{\mu}_{n0}$ & 0.2724 & 0.0320 & 0.0309 & 0.0301 &  & 0.2585 &0.0289 & 0.0280 & 0.0283 \\
$\hat{\mu}_{n1}$ & 0.3104 & 0.0318 & 0.0318 & 0.0318 &  & 0.3104 &0.0318 & 0.0318 & 0.0318 \\
$\hat{\mu}_{n1}-\hat{\mu}_{n0}$ & 0.0380 & 0.0451 & 0.0443 & 0.0438&  & 0.0519 & 0.0430 & 0.0423 & 0.0425\\
$\hat{\mu}_{c0}$ & 0.0351 & 0.1233 & 0.1233 & 0.1243 &  & 0.0685 & 0.1227 & 0.1160 & 0.1241 \\
$\hat{\mu}_{c1}$ & -0.0385 & 0.1398 & 0.1445 & 0.1507 &  & -0.1888& 0.0564 & 0.0564 & 0.0747 \\
$\hat{\mu}_{c1}-\hat{\mu}_{c0}$ & -0.0736 & 0.1909 & 0.1921 & 0.1951 &  & -0.2573 & 0.1339 &0.1305 & 0.1476\\
$\hat{\sigma}_{a0}$ & 0.2874 & 0.0082 & 0.0100 & 0.0121 &  & 0.2874 & 0.0082 & 0.0100 & 0.0121 \\
$\hat{\sigma}_{a1}$ & 0.2472 & 0.0124 & 0.0128 & 0.0134 &  & 0.2796 & 0.0112 & 0.0117 & 0.0132  \\
$\hat{\sigma}_{n0}$ & 0.2288 & 0.0237 & 0.0268 & 0.0305 &  & 0.2409 & 0.0214 & 0.0197 & 0.0196  \\
$\hat{\sigma}_{n1}$ & 0.3034 & 0.0198 & 0.0225 & 0.0255 &  & 0.3034 & 0.0198 & 0.0225 & 0.0255 \\
$\hat{\sigma}_{c0}$ & 0.4621 & 0.0856 & 0.0745 & 0.0652 &  & 0.4726 & 0.0900 & 0.0783 & 0.0695\\
$\hat{\sigma}_{c1}$ & 0.4551 & 0.1056 & 0.0892 & 0.0774 &  & 0.1138 & 0.0450 & 0.0309 & 0.0236\\
\hline\hline
\end{tabular}
\caption{Efficient Likelihood Estimates and Standard Errors from the univariate and the bivariate models applied to World War II data.} \label{tab:WWII}
\end{center}
\end{table}

\section{Conclusion}
We propose to sharpen the inference for a lower-dimensional mixture model by jointly
modeling the primary variable and an auxiliary variable. We have
established analytical results for several special cases that show
that the probability of correctly allocating mixture memberships
and the information number for the means of the primary outcome in
a bivariate mixture model with two Gaussian components are
generally larger than those in the corresponding univariate model.
The improvement under more general settings, including
misspecified models, is also observed in a comprehensive
simulation study and in two real data analyses. As shown in the second empirical example, there is in general no need to include many auxiliary variables, as most of the information gain comes from the auxiliary variable with a high association with the mixture membership.

The formal results we have obtained can be useful in many
settings, such as those mentioned in the introduction. The goal in empirical analysis, e.g.
in causal inference and small area estimation, should be to pick the \textit{best} auxiliary variable
that increases precision without increasing the risk of mis-specification. This issue will be the
subject of our future investigations.

\section*{Acknowledgements}
Mercatanti's research was partially supported by the U.S. NSF under Grant DMS-1127914 to the Statistical and Applied Mathematical Sciences Institute (SAMSI). Li's research was partially funded by the U.S.
NSF SES grants 1155697 and 1424688.

\section*{Appendix}

\noindent \textbf{Appendix 1}. \emph{Proof of Result 1.}
\begin{enumerate}
\item[(1)] Solving the equation $\partial \bd'\bV^{-1}\bd/ \partial d_2 =(2\sigma _{1}^{2}d_{2}-2\rho \sigma _{1}\sigma _{2}d_{1})~/~(\sigma
_{1}^{2}~\sigma _{2}^{2}-\rho ^{2}\sigma _{1}^{2}\sigma _{2}^{2})=0$, it is straightforward to show that $\underset{d_2}{\arg \min} \bd'\bV^{-1}\bd = \rho d_1\sigma_2/\sigma_1$, which gives the minimum of $\bd'\bV^{-1}\bd= d_1^2/\sigma_1^2$. \qed

\item[(2)] For equation $\partial \bd'\bV^{-1}\bd/ \partial \sigma_{12} =\partial \lbrack (d_{1}^{2}\sigma _{2}^{2}+d_{2}^{2}\sigma _{1}^{2}-2\sigma
_{12}d_{1}d_{2})~/~(\sigma _{1}^{2}~\sigma _{2}^{2}-\sigma
_{12}^{2})]/\partial \sigma _{12}= 0$, we have two solutions:
$\sigma _{12}=\sigma _{1}^{2}d_{2}/d_{1}$ or $\sigma _{2}^{2}d_{1}/d_{2}$. These two cannot hold at the same time due to the $\sigma_{12}\leq \sigma_1\sigma_2$.

When $\sigma _{12}=\sigma _{1}^{2}d_{2}/d_{1}$, $\rho=\sigma_{1}d_{2}/\sigma_2d_{1}$ and $\bd'\bV^{-1}\bd=d_1^2/\sigma_1^2$, which is the same value in the univariate case. When $\sigma _{12}=\sigma _{2}^{2}d_{1}/d_{2}$, $\rho=\sigma_{2}d_{1}/\sigma_1d_{2}$ and we can show
\[\bd'\bV^{-1}\bd=d_2^2/\sigma_2^2 \geq d_1^2/\sigma_1^2,\]
where the inequality is due to the constraint of $\sigma_{12}\leq \sigma_1\sigma_2$. \qed

\item[(3)]
Let $d_2=d_1\sigma_{12}/\sigma_1^2+\kappa$, with $\kappa \neq 0$, then we can show
\[\bd'\bV^{-1}\bd=\frac{d_{1}^{2}}{\sigma _{1}^{2}}+\frac{\kappa^{2}\sigma _{1}^{2}}{\sigma
_{1}^{2}~\sigma _{2}^{2}-\sigma _{12}^{2}}>\frac{d_{1}^{2}}{\sigma _{1}^{2}}.\]
Consequently, when $\kappa {\rightarrow} \infty$ ($\left\vert d_{2}\right\vert \rightarrow \infty$): $\bd'\bV^{-1}\bd\rightarrow \infty$, $\Pr(h(\bx_{i})<0 \mid k=1){\rightarrow} 1$, $\Pr (h(\bx_{i})>0 \mid k=1){\rightarrow} 0$, $\Pr (h(\bx_{i})>0 \mid k=2){\rightarrow} 1$, and $\Pr (h(\bx_{i})<0 \mid k=2){\rightarrow} 0$.
Moreover, it is straightforward to show that $\lim_{\left\vert d_{2}\right\vert \rightarrow \infty }f_{\bar{k}}(\bx_{i})=0$ for any $\bx_{i}$ lying on the discriminant line $h(\bx_{i})=0$ so that, given the local monotonicity of $f_{\bar{k}}(\bx_{i})$, we have:
 \[\lim_{\left\vert d_{2}\right\vert \rightarrow \infty }\{c~f_{k}(\mathbf{x}_{i})/f(\mathbf{x}_{i})\mid ({i\in k})\}=1.\]
The same arguments applies to prove
    \[\lim_{\left\vert \rho\right\vert \rightarrow 1 }\{c~f_{k}(\mathbf{x}_{i})/f(\mathbf{x}_{i})\mid ({i\in k})\}=1.\]
    \qed
\end{enumerate}

\bigskip
\noindent \textbf{Appendix 2}. \emph{Proof of Result 2.}

The diagonal terms of $\mathcal{I}_{1}$ have the form:
\[
\sum_{i}^{n}~c^{2}~[f_{k}(\bx_{i})/f(\bx_{i})]^{2}~\bV^{-1}(\bx_{i}-\bmu_{k})'(\bx_{i}-\bmu_{k})\bV^{-1}
\]
while the off-diagonals have:
\[
\sum_{i}^{n}~c~[f_{k}(\bx_{i})/f(\bx_{i})]~\bV^{-1}(\bx_{i}-\bmu_{k})'(\bx_{i}-\bmu_{\bar{k}})\bV^{-1}~\bar{c}~[f_{\bar{k}}(\bx_{i})/f(\bx_{i})]
\]
Given $\lim_{\left\vert d_{2}\right\vert \rightarrow \infty }\{c~f_{k}(\mathbf{x}_{i})/f(\mathbf{x}_{i})\mid ({i\in k})\}=1$ and $\lim_{\left\vert d_{2}\right\vert \rightarrow \infty }\{\bar{c}~f_{\bar{k}}(\mathbf{x}_{i})/f(\mathbf{x}_{i})\mid ({i\in k})\}=0$ we immediately have:
\[
\lim_{|d_2|\rightarrow \infty}\mathcal{I}_1=
\left(
\begin{array}{cc}
\mathbf{V}^{-1}\sum_{i\in k=1}~(\mathbf{x}_{i}-\hat{\bmu}_{1})(%
\mathbf{x}_{i}-\hat{\bmu}_{1})^{\prime }\mathbf{V}^{-1} & \mathbf{0}
\\
\mathbf{0} & \mathbf{V}^{-1}\sum_{i\in k=2}~(\mathbf{x}_{i}-\mathbf{\hat{\mu}%
}_{2})(\mathbf{x}_{i}-\hat{\bmu}_{2})^{\prime }\mathbf{V}^{-1}%
\end{array}
\right).
\]
The same arguments apply to prove the case when $\rho\rightarrow 1$, for fixed $(d_1, d_2, \sigma _{1}, \sigma _{2})$ and fixed sample size $n$.
%\[
%\lim_{\left\vert \rho\right\vert \rightarrow 1}\mathcal{I}_1=
%\left(
%\begin{array}{cc}
%\mathbf{V}^{-1}\sum_{i\in k=1}~(\mathbf{x}_{i}-\hat{\bmu}_{1})(%
%\mathbf{x}_{i}-\hat{\bmu}_{1})^{\prime }\mathbf{V}^{-1} & \mathbf{0}
%\\
%\mathbf{0} & \mathbf{V}^{-1}\sum_{i\in k=2}~(\mathbf{x}_{i}-\mathbf{\hat{\mu}%
%}_{2})(\mathbf{x}_{i}-\hat{\bmu}_{2})^{\prime }\mathbf{V}^{-1}%
%\end{array}
%\right).
%\]
\qed

\bigskip

\noindent \textbf{Appendix 3}. \emph{Proof of Result 3.}
\begin{enumerate}
\item[(1)] Let
\[
\frac{c~f_{k}(\mathbf{x}_{i})}{f(\mathbf{x}_{i})}=\left[ 1+\frac{\bar{c}~f_{\bar{k}}(x_{i1})\medskip (\sigma _{2\bar{k}}^{2})^{-1/2}\exp \left\{
-(x_{i2}-d_{2})^{2}/(2\sigma _{2\bar{k}}^{2})\right\} }{c~f_{k}(x_{i1})\medskip (\sigma _{2k}^{2})^{-1/2}\exp \left\{ -~x_{i2}^{2}/(2\sigma _{2k}^{2})\right\} }\right] ^{-1}=\left\{ 1+\frac{\bar{c}~f_{\bar{k}}(x_{i1})\medskip
}{c~f_{k}(x_{i1})\medskip }h(x_{i2})\right\} ^{-1},
\]

Notice that $c~f_{k}(\mathbf{x}_{i})/f(\mathbf{x}_{i})$ increases when $h(x_{i2})<1$, or
equivalently when:
\[
\frac{x_{i2}^{2}(\sigma _{2\bar{k}}^{2}-\sigma _{2k}^{2})}{2~\sigma _{2\bar{k}}^{2}~\sigma _{2k}^{2}}+\frac{x_{i2}~d_{2}}{\sigma _{2\bar{k}}^{2}}+\log
\frac{\sigma _{2k}^{2}}{\sigma _{2\bar{k}}^{2}}-\frac{d_{2}^{2}}{2~\sigma _{2\bar{k}}^{2}}=ax_{i2}^{2}+b(d_{2})x_{i2}+c(d_{2})<0\text{.}
\]
It is easy to show the discriminant of the quadratic form is always positive.

If $\sigma _{2k}^{2}<\sigma _{2\bar{k}}^{2}$, then $a>0$, so that $h(x_{i2})<1$ when
\[
\frac{-b(d_{2})-\sqrt{b(d_{2})^{2}-4ac(d_{2})}}{2a}=x_{2\inf }<x_{i2}<x_{2\sup }=\frac{-b(d_{2})+\sqrt{b(d_{2})^{2}-4ac(d_{2})}}{2a},
\]
and
\[
\Pr \left\{ h(x_{i2})<1|(i\in k)\right\}=\int \one\left[x_{2\inf}<x_{i2}<x_{2\sup}\right]f_{k}(x_{i2})dx_{i2}=\Phi _{k}(x_{2\sup})-\Phi _{k}(x_{2\inf}).
\]
It is easy to prove that $\lim_{\left\vert d_{2}\right\vert \rightarrow \infty }x_{2\inf}=-\infty$ and $\lim_{\left\vert
d_{2}\right\vert \rightarrow \infty }x_{2\sup }=+\infty$.
Consequently
\[
\lim_{\left\vert d_{2}\right\vert \rightarrow \infty }\Pr \left\{
h(x_{i2})<1~|~(i\in k)\right\} =1\text{.}
\]
Moreover, given that $\lim_{d_{2}\rightarrow +\infty }f_{\bar{k}}(x_{2\sup })=0$ and $\lim_{d_{2}\rightarrow -\infty }f_{\bar{k}}(x_{2\inf})=0$,
and given the local monotonicity of $f_{\bar{k}}(x)$, we have:
\[
\lim_{\left\vert d_{2}\right\vert \rightarrow \infty }\left\{ c~f_{k}(%
\mathbf{x}_{i})/f(\mathbf{x}_{i})|~(i\in k)\right\} =1\text{.}
\]

If $\sigma _{2k}^{2}>\sigma _{2\bar{k}}^{2}$, then $a<0$, so that $h(x_{i2})<1$ when
\[
\frac{-b(d_{2})-\sqrt{b(d_{2})^{2}-4ac(d_{2})}}{2a}=x_{2\inf}<x_{i2} \text{ or } x_{i2}>x_{2\sup}=\frac{-b(d_{2})+\sqrt{b(d_{2})^{2}-4ac(d_{2})}}{2a},
\]
and
\[
\Pr \left\{ h(x_{i2})<1|i\in k\right\} =\int \one \left[x_{2\inf }>x_{i2}\text{ or }x_{i2}>x_{2\sup}\right]f_{k}(x_{i2})dx_{i2}=1-\Phi _{k}(x_{2\sup})+\Phi _{k}(x_{2\inf})
\]

Again, it is easy to prove that $\lim_{d_{2}\rightarrow +\infty }x_{2\inf}=+\infty$ and $\lim_{d_{2}\rightarrow
-\infty }x_{2\sup}=-\infty$. Therefore
\[
\lim_{\left\vert d_{2}\right\vert \rightarrow \infty }\Pr \left\{
h(x_{i2})<1~|~~(i\in k)\right\} =1.
\]
Given that $\lim_{d_{2}\rightarrow +\infty }f_{_{\bar{k}}}(x_{2\inf})=0$ and $\lim_{d_{2}\rightarrow -\infty }f_{_{\bar{k}}}(x_{2\sup})=0$, and given the local monotonocity of $f_{\bar{k}}(x)$, we have again:
\[
\lim_{\left\vert d_{2}\right\vert \rightarrow \infty }\left\{ c~f_{k}(%
\mathbf{x}_{i})/f(\mathbf{x}_{i})~|~~(i\in k)\right\} =1\text{.}
\]

The case $\sigma _{2k}=\sigma _{2\bar{k}}$ is trivial. \qed

\item[(2)] The same arguments to prove Result 2 apply here.
\end{enumerate}%

\bigskip
\noindent \textbf{Appendix 4}. \emph{Proof of Result 4.}

To simplify the proof, make the transformation as in \eqref{parametric set}, where $\rho=0$.

The information number for the mean of the primary variable in group $k=1$, $\mu _{11}$, from the univariate model has the form : \footnote{%
For the primary variable in group $k=2$, the proof can be analogously
developed.}
\[
\sum_{i}^{n}p^{2}~\left\{\frac{\hat{f}_{1}(x_{i1})}{\hat{f}(x_{i1})}\right\}^{2}~\frac{x_{i1}^{2}~%
}{\sigma _{1}^{4}}=\sum_{i}^{n} q(x_{i1})^2\frac{x_{i1}^{2}}{\sigma _{1}^{4}},
\]
where $q(x)=\left\{ 1+\frac{(1-p)~f_{2}(x)}{p~f_{1}(x)}%
\right\} ^{-1}=\left\{1+\frac{(1-p)}{p}\exp(-\frac{d_{1}^{2}}{2\sigma_1^2}+\frac{x d_{1}}{\sigma_1^2})\right\}^{-1}$.
%\[
%=\sum_{i\in k=1}+\sum_{i\in k=2}\left\{ 1+\frac{(1-p)~\hat{f}_{2}(x_{i1})}{%
%p~\hat{f}_{1}(x_{i1})}\right\} ^{-2}~\frac{x_{i1}^{2}~}{\sigma _{1}^{4}}.
%\]

For a large i.i.d. sample, given the consistency of the MLE of the means, $1/n$
times the information number tends to:
\[
I=p\cdot \int_{-\infty }^{+\infty }q(x_{1})^{2}\cdot \frac{x_{1}^{2}~}{%
\sigma _{1}^{4}}\cdot \exp (-\frac{x_{1}^{2}}{2\sigma _{1}^{2}}%
)~dx_{1}~+~(1-p)\cdot \int_{-\infty }^{+\infty }q(x_{1})^{2}\cdot \frac{%
x_{1}^{2}~}{\sigma _{1}^{4}}\cdot \exp \left\{-\frac{(x_{1}-d_{1})^{2}}{2\sigma _{1}^{2}}\right\}~dx_{1},
\]
which can be simplified as follows:
\begin{eqnarray*}
I&=&\int_{-\infty }^{+\infty }p q(x_{1})^{2}\frac{x_{1}^{2}~}{\sigma_{1}^{4}}
\exp (-\frac{x_1^2}{2\sigma_{1}^{2}})\left\{1+\frac{1-p}{p}\exp (\frac{d_{1}^{2}}{2\sigma_{1}^{2}}+\frac{x_{1}d_{1}}{\sigma_{1}^{2}})\right\}
\,dx_{1} \\
&=&\int_{-\infty }^{+\infty }\frac{\frac{x_{1}^{2}~}{\sigma_{1}^{4}}\exp(-\frac{x_{1}^2}{2\sigma_{1}^{2}})}
{\left\{p+(1-p)\exp(-\frac{d_{1}^{2}}{2\sigma_1^2}+\frac{x_1 d_{1}}{\sigma_1^2})\right\}}
\,dx_{1}\text{.}
\end{eqnarray*}

Analogously, from a bivariate model, $1/n$ times the information number for the mean of the primary
variable in group $k=1$  given a large i.i.d. sample
tends to:
\begin{eqnarray*}
I\!I&=&p\cdot \int_{-\infty }^{+\infty }\int_{-\infty }^{+\infty
} q(x_{1},x_{2})^{2}\cdot \frac{x_{1}^{2}}{\sigma _{1}^{4}}\cdot \exp \left(-\frac{x_{1}^{2}}{2\sigma_{1}^{2}}-\frac{x_{2}^{2}}{2\sigma_{2}^{2}}\right)~dx_{1}~dx_{2}\\
&&~ + ~(1-p)\cdot \int_{-\infty }^{+\infty }\int_{-\infty }^{+\infty}q(x_{1},x_{2})^{2}\cdot \frac{x_{1}^{2}}{\sigma _{1}^{4}}\cdot \exp \left\{-\frac{(x_{1}-d_{1})^{2}}{2\sigma_{1}^{2}}-\frac{(x_{2}-d_{2})^{2}}{2\sigma _{2}^{2}}\right\}~~dx_{1}~dx_{2},
\end{eqnarray*}
where $q(x_{1},x_{2})=\left\{1+\frac{(1-p)}{p} \frac{f_{2}(x_{1},x_{2})}{f_{1}(x_{1},x_{2})}\right\}^{-1}=\left\{1+\frac{1-p}{p}
\exp(-\frac{d_{1}^{2}}{2\sigma_{1}^{2}}+\frac{x_{1}d_{1}}{\sigma_{1}^{2}}
-\frac{d_{2}^{2}}{2\sigma_{2}^{2}}+\frac{x_{2}d_{2}}{\sigma_{2}^{2}})\right\}^{-1}$.

The term $I\!I$ can be simplified as follows:
\begin{small}
\begin{eqnarray*}
I\!I&=&\int_{-\infty }^{+\infty }\!\!\!\int_{-\infty }^{+\infty }
p q(x_{1},x_{2})^{2}\frac{x_{1}^{2}~}{\sigma _{1}^{4}}
\exp \left(-\frac{x_{1}^{2}}{2\sigma_{1}^{2}}-\frac{x_{2}^{2}}{2\sigma_{2}^{2}}\right)
\left\{1+\frac{1-p}{p}
\exp(-\frac{d_{1}^{2}}{2\sigma_{1}^{2}}+\frac{x_{1}d_{1}}{\sigma_{1}^{2}}
-\frac{d_{2}^{2}}{2\sigma_{2}^{2}}+\frac{x_{2}d_{2}}{\sigma_{2}^{2}})\right\} \,dx_{1}\,dx_{2}\\
&=&\int_{-\infty }^{+\infty }\!\!\!\int_{-\infty }^{+\infty }p q(x_{1},x_{2})\frac{x_{1}^{2}~}{\sigma _{1}^{4}}
\exp \left(-\frac{x_{1}^{2}}{2\sigma_{1}^{2}}-\frac{x_{2}^{2}}{2\sigma_{2}^{2}}\right) \,dx_{1}\,dx_{2}\\
&=&\int_{-\infty }^{+\infty }\exp (-\frac{x_{2}^{2}}{2\sigma _{2}^{2}})
\int_{-\infty }^{+\infty }\frac{\frac{x_{1}^{2}~}{\sigma_{1}^{4}}\exp (-\frac{x_{1}^{2}}{2\sigma _{1}^{2}})}
{\left\{p+(1-p)\exp (-\frac{d_{1}^{2}}{2\sigma _{1}^{2}}+\frac{x_{1}d_{1}}{\sigma _{1}^{2}})\exp
(-\frac{d_{2}^{2}}{2\sigma_{2}^{2}}+\frac{x_{2}d_{2}}{\sigma_{2}^{2}})\right\}}\,dx_{1}\,dx_{2}.
\end{eqnarray*}
\end{small}

Suppose the distance\ $d_{2}>0$\footnote{%
If $d_{2}<0$, the proof can be analogously developed.}. If $x_{2}<d_{2}/2$, we have that $\exp (-\frac{1}{2}\frac{d_{2}^{2}}{\sigma _{2}^{2}}%
+x_{2}\frac{d_{2}}{\sigma _{2}^{2}})<1$; then
\[\left\{p+(1-p)\exp (-\frac{1}{2}\frac{d_{1}^{2}}{\sigma _{1}^{2}}+x_{1}\frac{d_{1}}{\sigma _{1}^{2}})\right\}^{-1}
<\left\{p+(1-p)\exp (-\frac{1}{2}\frac{d_{1}^{2}}{\sigma _{1}^{2}}+x_{1}\frac{d_{1}}{\sigma _{1}^{2}})\exp (-\frac{1%
}{2}\frac{d_{2}^{2}}{\sigma _{2}^{2}}+x_{2}\frac{d_{2}}{\sigma _{2}^{2}})\right\}^{-1}.
\]
Given that
\[
\int_{-\infty }^{{d_{2}}/{2}}\exp (-\frac{1}{2}\frac{x_{2}^{2}}{\sigma
_{2}^{2}})\,dx_{2}\geqslant \int_{-\infty }^{0}\exp (-\frac{1}{2}\frac{%
x_{2}^{2}}{\sigma _{2}^{2}})\,dx_{2}=\frac{\sqrt{2\pi }\sigma _{2}}{2},
\]%
we have:
\[
I\!I=\int_{-\infty }^{\frac{d_{2}}{2}}+\int_{\frac{d_{2}}{2}}^{+\infty
}\left[ \exp (-\frac{1}{2}\frac{x_{2}^{2}}{\sigma _{_{2}}^{2}}%
)\,\int_{-\infty }^{+\infty }\frac{\frac{x_{1}^{2}~}{\sigma _{1}^{4}}\exp (-%
\frac{1}{2}\frac{x_{1}^{2}}{\sigma _{1}^{2}})}{\left\{ p+(1-p)\exp (-\frac{1}{%
2}\frac{d_{1}^{2}}{\sigma _{1}^{2}}+x_{1}\frac{d_{1}}{\sigma _{1}^{2}})\exp
(-\frac{1}{2}\frac{d_{2}^{2}}{\sigma _{_{2}}^{2}}+x_{2}\frac{d_{2}}{\sigma
_{_{2}}^{2}})\right\} }\,dx_{1}\right] \,dx_{2}
\]

\[
>\int_{-\infty }^{\frac{d_{2}}{2}}\exp (-\frac{1}{2}\frac{x_{2}^{2}}{\sigma
_{2}^{2}})\,\,dx_{2}\int_{-\infty }^{+\infty }\frac{\frac{x_{1}^{2}~}{\sigma
_{1}^{4}}\exp (-\frac{1}{2}\frac{x_{1}^{2}}{\sigma _{1}^{2}})}{\left\{
p+(1-p)\exp (-\frac{1}{2}\frac{d_{1}^{2}}{\sigma _{1}^{2}}+x_{1}\frac{d_{1}}{%
\sigma _{1}^{2}})\right\} }\,dx_{1}
\]%
Consequently $I\!I>\frac{\sqrt{2\pi }\sigma _{2}}{2}I$, so that $I\!I>I$ when $\sigma _{2}\geq
\frac{2}{\sqrt{2\pi}}$. Moreover, the property (easy to prove) $I\!I$ possesses to be monotonically decreasing in $\sigma _{2}$ guarantees
that $I\!I>I$ always holds.

\qed

%\textbf{Appendix 5}. \emph{Analytical Form of the Hessian Matrix for Model \eqref{mixture}.}
%The subscript $i$ is suppressed for simplicity in the following discussion. For bivariate densities, $\bq_{m}=(q_{m1}, q_{m2})'$, and
%\[\bQ_{mm'}=
%\left(
%\begin{array}{cc}
%\partial q_{m1}/\partial \mu_{m1} & \partial q_{m1}/\partial \mu_{m2}  \\
%\partial q_{m'2}/\partial \mu_{m'1} & \partial q_{m'2}/\partial \mu_{m'2}
%\end{array}
%\right),
%\]
%for the four combinations of $m, m'$: $(1,1), (1,2), (2,1), (2,2)$. %\comment{there are some problems in notation here}
%
%Writing the above expression out (omitting $\bx$ and $i$), we have
%\begin{equation*}
%q_{mk}=\frac{c}{1-\rho_k^2}\cdot \frac{f_k}{f}\cdot \tau_{mk},
%\end{equation*}
%where $\tau_{mk} = \frac{x_m-\mu_{mk}}{\sigma_{mk}^2}-\frac{\rho_k(x_{m'}-\mu_{m'k})}{\sigma_{1k}\sigma_{2k}}.$
%The entries for the Hessian matrix are ($m\neq m', k\neq k'$)
%\begin{eqnarray*}
%\frac{\partial q_{mk}}{\partial \mu_{mk}}&=&\frac{c}{(1-\rho_k^2)f^2}\left(\frac{\bar{c}f_1f_2 \tau_{mk}^2}{1-\rho_k^2} -\frac{f_k f}{\sigma_{mk}^2} \right), \\
%\frac{\partial q_{mk}}{\partial \mu_{m'k}} &=&\frac{c}{(1-\rho_k^2)f^2}\left(\frac{\bar{c}f_1f_2 \tau_{1k}\tau_{2k}}{1-\rho_k^2} +\frac{\rho_k f_k f}{\sigma_{1k}\sigma_{2k}} \right), \\
%\frac{\partial q_{mk}}{\partial \mu_{mk'}} &=&-\frac{p(1-p)f_1 f_2 \tau_{m1}\tau_{m2}}{(1-\rho_1^2)(1-\rho_2^2)f^2}, \\
%\frac{\partial q_{mk}}{\partial \mu_{m'k'}} &=&-\frac{p(1-p)f_1 f_2 \tau_{mk}\tau_{m'k'}}{(1-\rho_1^2)(1-\rho_2^2)f^2}.
%\end{eqnarray*}

\bibliographystyle{plainnat}
\bibliography{mixture}

\end{document}